\documentclass [a4paper,10pt]{article}      % Specifies the documentclass
\usepackage[T1]{fontenc}
\usepackage[english]{babel}
\input epsfig.sty
\begin{document}
\hoffset-5mm
\arraycolsep.15em
\def\kmsMpc{{\mbox{km}{\cdot}\mbox{s}^{-1}{\cdot} \mbox{Mpc}^{-1}}}
\def\arctanh{\mathop{\rm arctanh}\nolimits}

\def\a{\alpha}
\def\b{\beta}
\def\g{\gamma}
\def\d{\delta}
\def\e{\epsilon}

\def\edoc{\end{document}}

\def\hang{\hangindent\parindent}
\def\textindent#1{\indent\llap{#1\enspace}\ignorespaces}
\def\litem{\par\hang\textindent}
\def\subitem{\par\indent \hangindent2\parindent\textindent}

\title{Evolution of density perturbations in a~realistic universe}
\author{Marek~Demia{\'n}ski$^1$, Zdzis{\l}aw A.~Golda$^2$, Andrzej~Woszczyna$^2$\\
${}^1$ Institute of Theoretical Physics, University of Warsaw,\\
 Warsaw, Poland {\it mde@fuw.edu.pl}\\
$^2$ Astronomical Observatory, Jagellonian University,\\
Cracow, Poland {\it uowoszcz@cyf-kr.edu.pl}}

%\date{}

\maketitle
\begin{abstract}
Prompted by the recent more precise determination of the basic
cosmological parameters and growing evidence that the
matter-energy content of the universe is now dominated by dark
energy and dark matter we present the general solution of the
equation that describes the evolution of density perturbations
in the linear approximation. It turns out that as in the
standard CDM model the density perturbations grow very slowly
during the radiation dominated epoch and their amplitude
increases by a factor of about 4000 in the matter and later dark
energy dominated epoch of expansion of the universe.

\end{abstract}

\section{Introduction}

During the last decade the possibilities of observing the
universe have substantially increased. Larger and better
telescopes, more sensitive CCD cameras and high resolutions
spectrographs have been built. These advances enabled more
thorough investigation of the content of the universe and more
accurate determination of the basic cosmological parameters.
Recently undertaken project of measuring the Hubble constant and
the deceleration parameter by using the type Ia supernovae as
standard candles lead to surprising results. It turned out that
the distant supernovae appear to be fainter than expected and in
order to reconcile observations with theoretical expectations it
is necessary to introduce the cosmological constant or dark
energy as the dominant constituent of the present universe. The
origin and nature of the dark energy is not know and therefore
in what follows we assume that it can be associated with the
cosmological constant.

By now there is a solid evidence that large fraction of the total
mass of galaxies and clusters of galaxies is in the form of non
baryonic dark matter of unknown nature and composition. Existence
of dark matter has been also confirmed by independent studies of
the motion of galaxies and their spatial distribution and results
of high redshift galaxy surveys (for example, 2dF and SDSS).

Recently released data on the anisotropy of the cosmic microwave
background radiation by the WMAP team when combined with
observations of the type Ia supernovae at high redshifts and
observed properties of clusters of galaxies and galaxy surveys
are all consistently determining the basic cosmological
parameters leading to what some already call the Standard
Cosmological Model~\cite{Bah&Ost&Per&Ste}.

It is customary to express the density of different components of
the universe in terms of the critical density $\varrho_{\rm
crit}=\displaystyle{3H_{0}^{2}\over {8\pi G}}$, where $H_0$ is the
present value of the Hubble constant and $G$ is the gravitational
constant. So we have
    \begin{equation}
\Omega_{\rm m}={{8\pi G}\over 3H_{0}^{2}}\varrho_{\rm m},~
\Omega_{\rm r}={{8\pi G}\over 3H_{0}^{2}}\varrho_{\rm r},~
\Omega_{\rm k}=-{kc^{2}\over H_{0}^{2}a_{0}^{2}},~\mbox{and}~
\Omega_{\Lambda}={{\Lambda c^{2}}\over 3H_{0}^{2}},
    \label{eq:00}
    \end{equation}
where $\varrho_{\rm m}$ and $\varrho_{\rm r}$ are
correspondingly the present density of matter and radiation
including relativistic particles, $k$ is the curvature parameter
(+1, 0, $-$1), $a_0$ --- the present value of the scale factor,
and $\Lambda$
--- the cosmological constant (dark energy).

The fundamental parameters of the most popular cosmological
models consistent with all sets of observational data, can be
taken as
    \[
\Omega_{\rm m}h^{2}=0.134\pm 0.006 ,~\Omega_\Lambda = 0.73\pm
0.04,~ \Omega_{\rm k}\approx 0,
    \]
    \[
\Omega_{\rm r}h^{2}=2.47\times 10^{-5},~{\rm and}~H_0= 71\pm 4
~{\kmsMpc} = 100\mbox{$h$}~\kmsMpc\,.
    \]
It is interesting to note that $\Omega_{\rm B}h^{-2}=
0.023\pm0.001$, and it is much less than $\Omega_{\rm m}$ and
moreover only small fraction of baryons is luminous $\Omega_{\rm
lum}=(0.003\pm 0.001)h^{-1}$. It has been established that
neutrinos are not massless but they have a small rest mass of
the order of 0.02 eV and therefore if they decoupled from the
state of thermodynamic equilibrium while relativistic their
contribution to the density of relativistic particles is
$\Omega_{\nu}h^{2}=\displaystyle{{\sum m_{i}}\over {94~{\rm
eV}}}$, where $m_{i}$ is the mass of $i$-th neutrino specie or
$\Omega_{\nu}\sim 6\times 10^{-4}$.

Prompted by these new results and in particular by the growing
evidence that the cosmological constant (dark energy) is different from zero we want
to analyze evolution of density perturbations in a model universe
filled in with matter, radiation and the cosmological constant.

\section{Evolution of density perturbations}

We assume that the background universe is homogeneous and isotropic
and its geometry is described by the Friedman--Robertson--Walker line
element
    \begin{equation}
ds^{2}=dt^{2}-a^{2}(t)\left[{dr^{2}\over {1-kr^{2}}}+
r^{2}(d\theta^{2}+\sin^2{\theta}d\varphi^{2})\right],
    \label{eq:1}
    \end{equation}
where $a(t)$ is the scale factor, $k=+1,~ 0,~ -1$, and we use
units such that the velocity of light $c=1$. The Einstein field
equations with the hydrodynamical energy-momentum tensor
describing the distribution of matter and the cosmological
constant $\Lambda$ reduce to:
    \begin{eqnarray}
    \label{eq:2}
2{{\ddot a}\over a}+\left({{\dot a}\over a}\right)^{2}+{k\over a^{2}}&=&
 -8\pi Gp+\Lambda,\\
\left({{\dot a}\over a}\right)^{2}+{k\over a^{2}}&=&
{{8\pi G}\over 3}\varrho+{1\over 3}\Lambda.
    \label{eq:3}
    \end{eqnarray}

We assume that the universe is filled in with non relativistic
matter (dust) and radiation (photons, neutrinos and other
relativistic particles) but the non relativistic matter does not
interact with radiation. We understand that this is a rather
drastic assumption and it restricts the applicability of our model
to the post decoupling period. The nature of dark matter particles
--- the dominant constituent of matter is not known. From
laboratory and astronomical observations it only follows that
they interact very weekly with baryons and photons and therefore
dark matter particles decoupled from the state of
thermodynamical equilibrium much earlier than baryons, quite
possibly at $z \gg 10^{6}$.

We can therefore decompose the total density of matter--energy into
dust and radiation and so
    \begin{equation}
\varrho=\varrho_{\rm r}+\varrho_{\rm m},
    \label{eq:4}
    \end{equation}
and
    \begin{equation}
p={1\over3}\varrho_{\rm r}.
    \label{eq:5}
    \end{equation}
In our model matter and radiation expand adiabatically and we have
    \begin{equation}
\varrho_{\rm r}a^{4}={\rm const},
    \label{eq:6}
    \end{equation}
and
    \begin{equation}
\varrho_{\rm m}a^{3}={\rm const}.
    \label{eq:7}
    \end{equation}

The Friedman equation (\ref{eq:3}) supplemented by (\ref{eq:4}),
(\ref{eq:5}), (\ref{eq:6}) and (\ref{eq:7}) form a
closed system of equations. Unfortunately when $\varrho_{\rm r}$,
$\varrho_{\rm m}$, and $\Lambda$ are all different from zero the Friedman
equation can not be solved analytically. However using the standard
notation $H= \displaystyle{{\dot a}\over a}$ and introducing the redshift
parameter $1+z=\displaystyle{a_{0}\over a}$ equation (\ref{eq:3}) can be
transformed into
    \begin{equation}
H^{2}(z)=
H^{2}_{0}\left(\Omega_{\rm r}(1+z)^{4}+\Omega_{\rm m}(1+z)^{3}+\Omega_{\rm k}(1+z)^{2}+
\Omega_{\Lambda}\right).
    \label{eq:8}
    \end{equation}
It is now apparent that if $\Omega_{\rm r}\not=0$ then
sufficiently early as $z \rightarrow \infty$ the density of
radiation determines the expansion rate of the universe.

The problem of evolution of small density perturbations in the
Friedman universe was thoroughly discussed by Lifshitz in 1946.
The general relativistic equations describing evolution of small
perturbations have been extensively studied since then. Here we
are interested in evolution of density perturbations in the
pressureless (dust) component of matter. Hypersurfaces
orthogonal to the dust matter flow coincide with the equal
proper time hypersurfaces, therefore the perturbation equation
has the same form in both, the  Newtonian theory and
relativistic gauge-invariant
formalisms~\cite{Brand&Kahn&Press,Kodama&Sasaki,Ellis&Bruni,Padmanabhan,Peebles}
    \begin{equation}
{\ddot \Delta}+2H{\dot \Delta}-4\pi G\varrho_{\rm m}\Delta =0\,,
    \label{eq:9}
    \end{equation}
where by capital $\Delta$ we denote the density contrast
$\Delta=\displaystyle{{\delta \varrho}\over \varrho}$ measured
on the flow orthogonal hypersurfaces, and dot stands for the
derivative with respect to time $t$. Since the time dependence
of $H$ and $\varrho_{\rm m}$ is in the general case not known,
instead of time we will use another independent variable
$x=\displaystyle{a(t)\over a_{0}}$ so $x$ is changing from~0
to~1. Using the relation $\displaystyle{dx\over dt}=xH$ and
(\ref{eq:8}) we transform the equation (\ref{eq:9}) into
    \begin{equation}
    x(\Omega_{\rm r}+\Omega_{\rm m}x+\Omega_{\rm k}x^{2}+\Omega_{\Lambda}x^4)
{\Delta''}+(\Omega_{\rm r}+{3\over 2}\Omega_{\rm m}x+
2\Omega_{\rm k}x^{2}+3\Omega_{\Lambda}x^{4}){\Delta'}-
{3\over 2}\Omega_{\rm m}\Delta=0,
    \label{eq:10}
    \end{equation}
where the prime denotes differentiation with respect to $x$.

The differential equation~(\ref{eq:10}) has a very special
form, the coefficients of the second and first derivative are
polynomials of the independent variable $x$. In turn the
coefficients of these polynomials form a vector
$\Omega=\{\Omega_{\rm m},\Omega_{\rm r},\Omega_{\rm k}, \Omega_{\Lambda}\}$
in the 4--dimensional parameter space. The Friedman equation (\ref{eq:3})
constrains these parameters imposing the condition that
   \begin{equation}
\Omega_{\rm r}+\Omega_{\rm m}+\Omega_{\rm k}+\Omega_{\Lambda}=1.
    \label{eq:10a}
    \end{equation}

In the parameter space this relation defines a hyperplane. It
means that only three of the parameters are independent. The
polynomial multiplying the second derivative in the
equation~(\ref{eq:10}) determines the number and positions of
singular points of this equation in the finite domain. In the
generic case this equation has six regular singular points
including one at infinity. For certain values of $\Omega$s some
of the singular points became complex. To study the general
properties of this equation it is therefore natural to consider
 the independent variable as complex. In the complex
domain the equation~(\ref{eq:10}) is of  Fuchs type~\cite{Ince}
with six regular singular points. General solutions of such
differential equations are practically unknown. We will seek and
investigate solutions of this equation in two cases: when
$\Omega_{\Lambda}=0$ equation~(\ref{eq:10}) reduces to the Heune
equation~\cite{Heune,Ronveux,Whittaker&Watson,Snow} with four
regular singular points, and in the general case when
$\Omega_{\Lambda}\not=0$  the equation~(\ref{eq:10}) has six
regular singular points.

\section{Solutions with $\Omega_{\Lambda}=0$}

When $\Omega_{\Lambda}=0$ the equation~(\ref{eq:10}) assumes the
form
    \begin{equation}
x(\Omega_{\rm r}+\Omega_{\rm m}x+\Omega_{\rm
k}x^{2}) {\Delta''}+(\Omega_{\rm r}+{3\over 2}\Omega_{\rm m}x+
2\Omega_{\rm k}x^{2}){\Delta'}- {3\over 2}\Omega_{\rm m}\Delta=0\,.
    \label{eq:10b}
    \end{equation}
This equation has four regular singular points at
$x_1=0,~x_{2,3}=\displaystyle{\frac{-\Omega_{\rm
m}\mp\sqrt{\Omega_{\rm m}^2-4\Omega_{\rm k}\Omega_{\rm
r}}}{2\Omega_{\rm k}}}$ and~$x_4=\infty$. General discussion of
solutions of this equation in the parameter space $\{\Omega_{\rm
m},\Omega_{\rm r},\Omega_{\rm k}\}$ will be provided after we
present some of the particular solutions.

{\large$\bullet$} The case $\Omega=\{\Omega_{\rm m},0,0,0\}$.
Exact solutions of equation~(\ref{eq:10b}) are known for
particular values of the parameters $\Omega=\{\Omega_{\rm
m},\Omega_{\rm r},\Omega_{\rm k},\Omega_{\Lambda}\}$. When
$\Omega=\{\Omega_{\rm m},0,0,0\}$, the equation (\ref{eq:10}) can
be easily solved and we get
    \begin{equation}
\Delta=c_{1}\Delta_+ +c_{2}\Delta_-=c_{1}x+
c_{2}x^{-\frac32},
    \label{eq:11}
    \end{equation}
where by $c_{1}, c_{2}$ we denote arbitrary constants not only
here but throughout this paper. It is apparent that
$\Delta_{+}(x)$ does vanish at $x=0$ and increases with increasing
$x$. It describes the growing mode of density perturbations. The
other solution $\Delta_{-}(x)$ is singular at $x=0$ and decreases
with increasing $x$. It describes the decaying mode of density
perturbations. This solution was know already to Lifshitz in
1946~\cite{Lifshitz}.

{\large$\bullet$} The case $\Omega=\{\Omega_{\rm m},\Omega_{\rm r},0,0\}$.
Slightly more general solution of the equation~(\ref{eq:10b}) is
obtained when $\Omega=\{\Omega_{\rm m},\Omega_{\rm r},0,0\}$
\cite{Groth&Peebles,Meszaros}
    \begin{eqnarray}
\Delta&=&c_{1}\Delta_{+}+c_{2}\Delta_{-}=c_{1} \left(\Omega_{\rm
r}+{3\over 2}\Omega_{\rm m}x\right)\nonumber\\
&&{}+c_{2}{15\over 4}{\Omega_{\rm m}^{3\over2}\over  \Omega_{\rm
r}^{2}}\left(-3\sqrt{\Omega_{\rm r}+\Omega_{\rm m}x}-{1\over
\sqrt{\Omega_{\rm r}}}
\left(\Omega_{\rm r}+{3\over 2}\Omega_{\rm m}x\right)
\ln{{\sqrt{\Omega_{\rm r}+\Omega_{\rm m}x}- \sqrt{\Omega_{\rm
r}}}\over {\sqrt{\Omega_{\rm r}+\Omega_{\rm
m}x}+\sqrt{\Omega_{\rm r}}}}\right).
    \label{eq:12}
    \end{eqnarray}
As in the previous case $\Delta_{+}(x)$ describes the growing mode
of density perturbations, but now $\Delta_{+}(0)=\Omega_{\rm r}$.
The other independent solution $\Delta_{-}(x)$ is singular at
$x=0$ and decreases with increasing $x$. The normalization of
these solutions has been chosen in such a way that when
$\Omega_{\rm r}\rightarrow 0$, $\Delta_{+}$ and  $\Delta_{-}$ tend
to the corresponding modes of the solution (\ref{eq:11}).

{\large$\bullet$} The case $\Omega=\{\Omega_{\rm m},0,\Omega_{\rm k},0\}$.
Another exact solution is obtained when the cosmological constant
is zero and radiation is not present, in this case the non zero
parameters are $\Omega=\{\Omega_{\rm m},0,\Omega_{\rm k},0\}$ and
the solution can be written in the form~\cite{Weinberg,Peebles}
    \begin{eqnarray}
\Delta&=&c_{1}\Delta_{+}+c_{2}\Delta_{-}\nonumber\\
&=&c_{1}\left[\frac{5}{2}\frac{\Omega_{\rm m}}{\Omega_{\rm k}}
+\frac{15}{2}\frac{\Omega_{\rm m}^2}{\Omega_{\rm k}^2}\frac{1}{x}
    -\frac{15}{2}\frac{\Omega_{\rm m}^{5\over2}}{\Omega_{\rm
k}^{5\over2}} {x^{-\frac32}}\sqrt{1+ \frac{\Omega_{\rm
k}}{\Omega_{\rm m}}x}\ln \left(\sqrt{\frac{\Omega_{\rm
k}}{\Omega_{\rm m}}x}+\sqrt{1+ \frac{\Omega_{\rm k}}{\Omega_{\rm
m}}x}
\right)\right]\nonumber\\
&&{}+c_2\sqrt{1+\frac{\Omega_{\rm k}}{\Omega_{\rm
m}}x}\,x^{-\frac{3}{2}}.
    \label{eq:12a}
    \end{eqnarray}
As before  $\Delta_{+}(x)$ describes the growing mode of density
perturbations, in this case $\Delta_{+}(0)=0$ and asymptotically
for large $x$ it tends to $\lim\limits_{x\to
\infty}\Delta_{+}(x)=\displaystyle{\frac{5}{2}\frac{\Omega_{\rm
m}}{\Omega_{\rm k}}}$. The second independent solution
$\Delta_{-}(x)$ is singular at~$x=0$ and it decreases to zero
when $x\rightarrow \infty$. When $\Omega_{\rm k}\rightarrow 0$
then both modes of density perturbations $\Delta_{+}$ and
$\Delta_{-}$ tend to the corresponding modes of the solution
(\ref{eq:11}).

{\large$\bullet$} The case $\Omega=\{2\sqrt{\Omega_{\rm
r}}(1-\sqrt{\Omega_{\rm r}}),\Omega_{\rm r},\Omega_{\rm k},0\}$.
Another particular solution of the evolution equation
(\ref{eq:10b}) in a form of first several terms of a series
expansion has been given by Guyot and
Zeldovich~\cite{Guyot&Zeldovich}. This solution corresponds to
the following choice of the parameters
$\Omega=\{2\sqrt{\Omega_{\rm r}}(1-\sqrt{\Omega_{\rm
r}}),\Omega_{\rm r},\Omega_{\rm k},0\}$ and for this choice of
parameters two singular points of the equation (\ref{eq:10b})
$x_2=x_3$ overlap. In this case the equation (\ref{eq:10b})
assumes the form
    \begin{equation}
x(2\Omega_{\rm r}+\Omega_{\rm m}x)^2 \Delta''+ 2(\Omega_{\rm
r}+\Omega_{\rm m}x)(2\Omega_{\rm r}+\Omega_{\rm m}x)\Delta' -
6\Omega_{\rm r}\Omega_{\rm m}\Delta=0.
    \label{eq:20}
    \end{equation}
This equation can be solved exactly and the exact solution is
expressed by hypergeometric functions
    \begin{eqnarray}
\Delta&=&c_{1}\,\Delta_{+}+c_{2}\Delta_{-}\nonumber\\
&=&c_{1}\,{\Omega_{\rm r}}\;{}_2F_1
    \left(\sqrt{3}\,i,-\sqrt{3}\;i,
1,\frac{\Omega_{\rm m}x}{2\Omega_{\rm r}+\Omega_{\rm m}x}
    \right)\nonumber\\
    &&+c_2
    \left[
-\frac{\sqrt3\pi}{\sinh[\sqrt3\pi]}
\frac{2\Omega_{\rm r}^2}{2\Omega_{\rm r}+\Omega_{\rm m}x}
\;{}_2F_1
    \left(1+\sqrt{3}\,i,1-\sqrt{3}\;i,
2,\frac{2\Omega_{\rm r}}{2\Omega_{\rm r}+\Omega_{\rm m}x}
    \right)
    \right].
    \label{eq:35}
    \end{eqnarray}
Despite the appearance of imaginary unit the solutions are real
because the first two arguments in the function ${}_2F_1$ are
complex conjugate of each other.  $\Delta_+$ describes the
growing mode with $\Delta_+(0)=\Omega_{\rm r}$ and
asymptotically for large $x$ we get $\lim\limits_{x\to\infty}
\Delta_+(x)=\displaystyle{ \Omega_{\rm
r}\frac{\sinh(\sqrt{3}\pi)}{\sqrt{3}\pi}}$. The second
independent solution represents the decaying mode, it is
singular at $x=0$ and when $x$ grows to infinity this solution
tends to zero.

\begin{figure}[h]
\centerline{\psfig{file=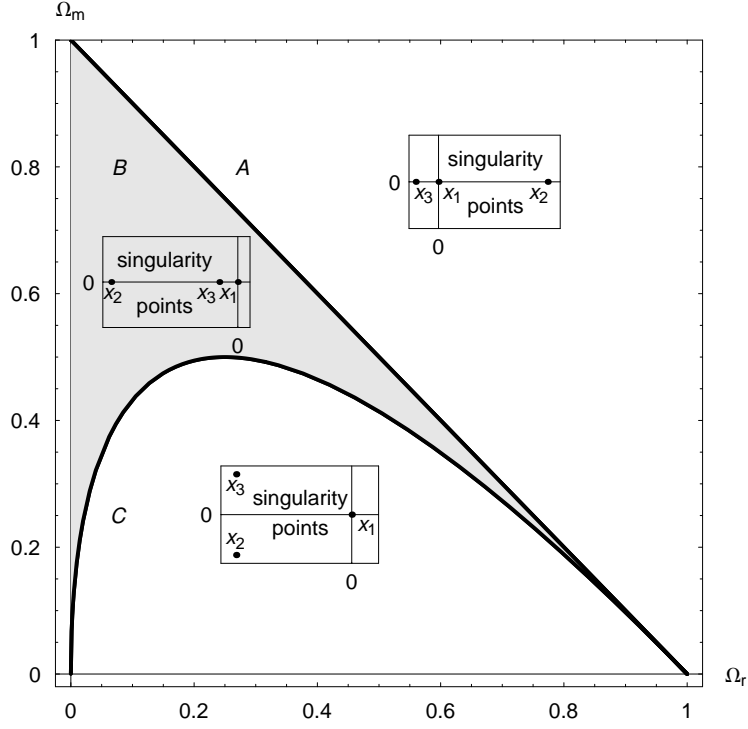,width=100mm,angle=0,clip=}}
\caption{The $\{\Omega_{\rm r},\Omega_{\rm m}\}$ plane with various domains and
singularity points.}
    \label{fig:1}
    \end{figure}

{\large$\bullet$} The case $\Omega=\{\Omega_{\rm m},\Omega_{\rm
r},\Omega_{\rm k},0\}$. Let us now return to the general
discussion. From the expressions determining the positions of
the regular singular points $x_2$ and~$x_3$ it is apparent that
in this case the parameter space is reduced to the plane of
$\{\Omega_{\rm r},\Omega_{\rm m}\}$. Physically interesting is
only a section of this plane described by $\Omega_{\rm r}\in
[0,1]$ and~$\Omega_{\rm m}\in (0,1]$ shown on Fig.~\ref{fig:1}.
On this figure there are
three regions: $A$, $B$ and  $C$. The boundary curve between the
regions  $B$ and $C$ represents the condition that two singular
points $x_2$ and $x_3$ overlap. This curve is described by the
equation $\Omega_{\rm m}^2-4\Omega_{\rm k}\Omega_{\rm r}=0$.
This equation combined with the constrain (\ref{eq:10a}) with
$\Omega_{\Lambda}=0$ leads to an explicit equation of this curve
$\Omega_{\rm m}=2\sqrt{\Omega_{\rm r}} (1-\sqrt{\Omega_{\rm
r}})$. Therefore the solution presented by Guyot
and~Zeldovich~\cite{Guyot&Zeldovich} is very special and in the
parameter space $\{\Omega_{\rm r},\Omega_{\rm m}\}$ it is
represented by a curve. The constrain (\ref{eq:10a}) with the
additional assumption that $\Omega_{\Lambda}=\Omega_{\rm k}=0$
reduces to an equation of a straight line $\Omega_{\rm
m}+\Omega_{\rm r}=1$, which is a~boundary between the regions
$A$ and $B$. Therefore  the regions $B$ and $C$ represent models
with~$\Omega_{\rm k}>0$ and the region $A$ represents models
with $\Omega_{\rm k}<0$. In what follows we will concentrate on
the region  $B$ (shadow region), as it is the most interesting
from the point of view of the present day observational data.
The general equation describing evolution of the density
perturbations (\ref{eq:10b}) in the region $B$ with
$\Omega=\{\Omega_{\rm m},\Omega_{\rm r},\Omega_{\rm k},0\}$ can
be turned into the Heune equation by the following
transformation of the independent variable
$\xi=\displaystyle{\frac{(\Omega_{\rm m}+\sqrt{\Omega_{\rm
m}^2-4\Omega_{\rm r}\Omega_{\rm k}})x} {2\Omega_{\rm
r}+(\Omega_{\rm m}+\sqrt{\Omega_{\rm m}^2-4\Omega_{\rm
r}\Omega_{\rm k}})x}}$, which shifts the positions of singular
points of equation (\ref{eq:10b}) form $\{x_1,x_2,x_3,x_4\}$ to
$\{0,d,\infty,1\}$, where $d$ is defined below. Then the
parameters of the Heune equation are given by:
$\alpha=\frac{1}{2}$, $\beta=0$, $\gamma=1$, $\delta=0$,
$\epsilon=\frac{1}{2}$,
$d=\displaystyle{\frac{1}{2}\left[1+\frac{\Omega_{\rm
m}}{\sqrt{\Omega_{\rm m}^2-4\Omega_{\rm r}\Omega_{\rm
k}}}\right]}>1$, $q=\displaystyle{\frac{3}{2}\frac{\Omega_{\rm
m}}{\sqrt{\Omega_{\rm m}^2-4\Omega_{\rm r}\Omega_{\rm k}}}}$ and
the equation assumes the canonical form of the Heune
equation~(\ref{eq:A01}). Solutions of this equation near the
singular point $\xi=0$ can be written in a form of
series~(\ref{eq:A04}) and (\ref{eq:A08})
    \begin{eqnarray}
\Delta=c_{1}\Delta_{+}+c_{2}\Delta_{-}
=c_{1}F(d,q;\alpha,\beta,\gamma,\delta;\xi)
+c_{2}G(d,q;\alpha,\beta,\gamma,\delta;\xi)\,,
    \label{eq:45}
    \end{eqnarray}
where $F$ and $G$ represent two independent solutions given in
Appendix A. As in the previous cases $\Delta_+$ is regular at
$\xi=0$ and represents the growing mode and~$\Delta_-$ is
singular at $\xi=0$ and describes the decaying mode. Solutions
of the equation (17) with the redshift parameter $z$ as an
independent variable in a form of a series expansion  have been
given by Rozgacheva and Sunyaev [17] but they do not discuss
convergence of these series.

\section{The general case when $\Omega_{\Lambda}\neq0$}

{\large$\bullet$} The case $\Omega=\{\Omega_{\rm
m},0,0,\Omega_{\Lambda}\}$. Let us first consider some special
cases of particular values of the parameters such that exact
solutions of the evolution equation of density perturbations~(\ref{eq:10})
are known. We start with a~case when the universe
is flat and without radiation, so $\Omega=\{\Omega_{\rm
m},0,0,\Omega_{\Lambda}\}$. In this case the general solution of
the equation (\ref{eq:10}) can be written down in the form
    \begin{eqnarray}
\Delta=c_{1}\,\Delta_{+}+c_{2}\Delta_{-}
=c_{1}x\;{}_2F_1
    \left(
\frac{1}{3},1,\frac{11}{6},-
\frac{\Omega_{\Lambda}}{\Omega_{\rm m}}x^3
    \right)
+c_2\sqrt{1+\frac{\Omega_{\Lambda}}{\Omega_{\rm m}}x^3}\,x^{-\frac{3}{2}}
    \label{eq:30A}
    \end{eqnarray}
(see~\cite{Chernin&Nagirner&Starikova}). $\Delta_{+}(x)$
describes the growing mode of density perturbations, in this
case we have $\Delta_{+}(0)=0$ and $\Delta_{+}(x)$ is
monotonically growing and asymptotically it tends to
$\lim\limits_{x\to\infty}\Delta_{+}(x)=\displaystyle{
\frac{5}{6\sqrt{\pi}}\Gamma\left[\frac{2}{3}\right]\Gamma\left[\frac{5}{6}\right]
\left[\frac{\Omega_{\Lambda}}{\Omega_{\rm
m}}\right]^{-1\slash3}}$. The second independent solution
$\Delta_{-}(x)$ is singular at~$x=0$ and it decreases with
increasing~$x$ asymptotically tending to
$\lim\limits_{x\to\infty}\Delta_{-}(x)=
\displaystyle{\sqrt{\frac{\Omega_{\Lambda}}{\Omega_{\rm m}}}}$.
When $\Omega_{\Lambda}\rightarrow 0$ the solutions $\Delta_{+}$
and $\Delta_{-}$ tend to~(\ref{eq:11}). The solution describing
the growing mode has been found by Eisenstein~\cite{Eisenstein}
and written in the form of an elliptic integral, it turns out
that it can be written in terms of hypergeometric function what
is more convenient to handle. In ~\cite{Bildhauer&Buchert&Kasai}
the solution $\Delta_{+}(x)$ is written in terms of the
incomplete beta-function.

{\large$\bullet$} The case $\Omega=\{\Omega_{\rm
m},0,\Omega_{\rm k},\Omega_{\Lambda}\}$. When radiation is not
present the parameter space is spanned by $\Omega=\{\Omega_{\rm
m},0,\Omega_{\rm k},\Omega_{\Lambda}\}$. In this case the
general solution of equation (\ref{eq:10}) can be written down
in the form~\cite{Heath}
    \begin{eqnarray}
\Delta=c_{1}\,\Delta_{+}+c_{2}\Delta_{-}
&=&c_{1} \sqrt{1+\frac{\Omega_{\rm k}}{\Omega_{\rm m}}x+\frac{\Omega_{\Lambda}}{\Omega_{\rm m}}x^3}\;x^{-\frac{3}{2}}
\int\frac{x^{3\slash 2}}{(\Omega_{\rm m}+\Omega_{\rm k}x+\Omega_{\Lambda}x^3)^{3\slash 2}}dx\nonumber\\
&&\null+c_2\sqrt{1+\frac{\Omega_{\rm k}}{\Omega_{\rm m}}x+\frac{\Omega_{\Lambda}}{\Omega_{\rm m}}x^3}\;x^{-\frac{3}{2}}.
    \label{eq:30B}
    \end{eqnarray}
The integral appearing in (\ref{eq:30B}) can be expressed in
terms of the Jacobi elliptic functions and its explicit form is
given in Appendix~C.

{\large$\bullet$} The case $\Omega=\{\Omega_{\rm m},\Omega_{\rm
r},\Omega_{\rm k},\Omega_{\Lambda}\}$. The most general
case, when  $\Omega=\{\Omega_{\rm m},\Omega_{\rm r},\Omega_{\rm
k},\Omega_{\Lambda}\}$, includes many cosmological models
(see Appendix D) and therefore in what follows we will limit our
discussion to the case when $\Omega_{\Lambda}>0$. From the weak
energy condition for matter and radiation it follows that
$\Omega_{\rm m}+\Omega_{\rm r}\geq 0$ what with
$\Omega_{\Lambda}\geq0$ and $\Omega_{\rm k}\geq0$ restricts the
parameter space to a cube with a size length equal to
one~\cite{Goliath&Ellis}
    \begin{equation}
0\leq\Omega_{\rm m}+\Omega_{\rm r}\leq 1,~~0\leq\Omega_{\rm k}\leq 1,~~0\leq\Omega_{\Lambda}\leq 1.
    \label{eq:40}
    \end{equation}
This cube then represents only models of the type~(H) in
Fig.~\ref{fig:2} (see also Appendix D). In this general case
the evolution equation~(\ref{eq:10}) is of Fuchs type with six
singular points. Positions of the proper singular points are
determined by the polynomial in front of ~$\Delta''(x)$, so we
have
    \begin{equation}
x(\Omega_{\rm r}+\Omega_{\rm m}x+\Omega_{\rm k}x^{2}+\Omega_{\Lambda}x^4)=0.
    \label{eq:41}
    \end{equation}
From the imposed conditions~(\ref{eq:40}) it follows that the
equation~(\ref{eq:41}) does not posses positive roots. It is
apparent that one root is  $x_1=0$. The other roots are either
all complex or two are complex and two are negative. Using the
standard technique of solving 4-th order algebraic
equations~\cite{Sierpinski} we find that
    \begin{eqnarray*}
x_1&=&0,\\
x_2&=&\frac{1}{2}(u_1+v_1+w_1),\\
x_3&=&\frac{1}{2}(-u_1-v_1+w_1),\\
x_4&=&\frac{1}{2}(u_1-v_1-w_1),\\
x_5&=&\frac{1}{2}(-u_1+v_1-w_1),\\
x_6&=&\infty,
%   \label{eq:42}
    \end{eqnarray*}
where we have added the singular point at infinity and
    \begin{eqnarray*}
u_1&=&\left[\frac{\sqrt{a^2+b^2}+a}{2}\right]^{1/2}+{\tt i}\left[\frac{\sqrt{a^2+b^2}-a}{2}\right]^{1/2},\\
v_1&=&\left[\frac{\sqrt{a^2+b^2}+a}{2}\right]^{1/2}-{\tt i}\left[\frac{\sqrt{a^2+b^2}-a}{2}\right]^{1/2}, \\
w_1&=&-\sqrt{-2a-2\frac{\Omega_{\rm k}}{\Omega_{\Lambda}}}, \\
a&=&-\frac{1}{2}\left\{\left[-q/2+\sqrt{\cal D}\right]^{1/3}
+\left[-q/2-\sqrt{\cal D}\right]^{1/3}\right\}-\frac{2}{3}\frac{\Omega_{\rm k}}{\Omega_{\Lambda}},\\
b&=&-\frac{1}{2}\left\{\left[-q/2+\sqrt{\cal D}\right]^{1/3}
-\left[-q/2-\sqrt{\cal D}\right]^{1/3}\right\},
    \end{eqnarray*}
while  ${\cal D}=q^2/4+p^3/{27}$,
$p=-\displaystyle{\frac{\Omega_{\rm k}^2+12\Omega_{\rm
r}\Omega_{\Lambda}}{3\Omega_{\Lambda}^2}}$,
$q=-\displaystyle{\frac{2\Omega_{\rm k}^3+27\Omega_{\rm
m}^2\Omega_{\Lambda}- 72\Omega_{\rm r}\Omega_{\rm
k}\Omega_{\Lambda}}{27\Omega_{\Lambda}^3}}$. We ordered the
roots in such a way that $x_2$ is the closest one to $x_1$. With
such order of the roots the homographic transformation
$\xi=\displaystyle{\frac{x}{x-x_2}}$ moves the singular points
$\{x_1,x_2,x_3,x_4,x_5,x_6\}$ of the equation~(\ref{eq:10}) into
$\{0,\infty,d_1,d_2,d_3,1\}$, where
$d_1=\displaystyle{\frac{x_3}{x_3-x_2}}$,
$d_2=\displaystyle{\frac{x_4}{x_4-x_2}}$, and
$d_3=\displaystyle{\frac{x_5}{x_5-x_2}}$. This transformation
maps the interval $x\in (0,\infty)$ into $\xi\in (0,1)$. Under
this transformation the equation~(\ref{eq:10}) assumes its
canonical form~(\ref{eq:B01}) with the following values of the
parameters $\a=\frac12$, $\b=0$, $\g=1$, $\d=-1$,
$\e_1=\e_2=\e_3=\frac12$. The additional parameters are given by
$q=-q_0$, $q_1=-2q_0$, $q_2=q_0$ where
$q_0=\displaystyle{\frac{3\Omega_{\rm m}x_2}{8\Omega_{\rm
r}+6\Omega_{\rm m} x_2+4\Omega_{\rm k}x_2^2}}$. The general solution
of the equation~(\ref{eq:10}) near the singular point
$\xi=0$ can be expressed in terms of the series~(\ref{eq:B05}) and
(\ref{eq:B10}) (see Appendix~B) and we have
    \begin{eqnarray}
\Delta=c_{1}\Delta_{+}+c_{2}\Delta_{-}
=c_{1}F(d_1,d_2,d_3,q,q_1,q_2;\alpha,\beta,\gamma,\delta,\epsilon_1,\epsilon_2;\xi)
+c_{2}G(d_1,d_2,d_3,q,q_1,q_2;\alpha,\beta,1,\delta,\epsilon_1,\epsilon_2;\xi).\nonumber\\
    \label{eq:50}
    \end{eqnarray}
The solution~$\Delta_+$ represents the growing mode and
$\Delta_{+}(0)\not=0$. The solution $\Delta_-$ represents
the decreasing mode which is singular at $\xi=0$.

\section*{Conclusions}

The general solutions of the equation that governs evolution of
density perturbations derived and discussed above can be used to
estimate the amplification factor of their amplitude.
Unfortunately very little is known about properties of the dark
matter particles and their interactions with the baryonic
matter. We will assume that dark matter particles decoupled from
the state of thermodynamical equilibrium at $z\approx 10^{8}$.
This is quite conservative estimate but our general conclusions
only weakly depend on this assumption. From the present values
of radiation energy density and matter density $\Omega_{\rm
r}\approx 5{\cdot}10^{-5}$ and $\Omega_{\rm m}\approx 0.7$ it
follows that until about $z\approx 10^{4}$ the universe was
radiation dominated and from that moment to about $z\approx 0.4$
it was matter dominated and only recently its expansion rate
became dark energy dominated. Using eq.~(\ref{eq:50}) we
calculated the amplification factor of the amplitude of density
perturbations in a universe with $\Omega_{\rm m}=0.3$,
$\Omega_{\Lambda}=0.7$, $\Omega_{\rm r}=5{\cdot}10^{-5}$ and we
got that $\Delta_{+}(z=10^{4})/\Delta_{+}(z=10^{8})\approx 1.9$,
$\Delta_{+}(z=0)/\Delta_{+}(z=10^{4})\approx 3691$ and
$\Delta_{+}(z=0)/\Delta_{+}(z=10^{8})\approx 7012$. Using
eq.~(\ref{eq:50}) we also calculated this amplification factor
for an open CDM model with $\Omega_{\rm m}=0.3$, $\Omega_{\rm
r}=5{\cdot} 10^{-5}$ and $\Omega_{\rm k}=0.7$ and we obtained
that $\Delta_{+}(z=10^{4})/\Delta_{+}(z=10^{8})\approx 1.9$,
$\Delta_{+}(z=0)/\Delta_{+}(z=10^{4})\approx 2170$ and
$\Delta_{+}(z=0)/\Delta_{+}(z=10^{8})\approx 4122$. This result
confirms the well known fact (see~\cite{Padmanabhan,Peebles})
that when the expansion rate of the universe is determined by
the energy of radiation and relativistic particles the density
perturbations grow very slowly (see eq.~(\ref{eq:12})). The
amplitude of density perturbations grows by a factor of about
few times $10^{3}$ in subsequent epochs. This puts strong
constrains on the process of generation of the primordial
density perturbations. The initial amplitude of the primordial
density perturbations has to be sufficiently large to grow to
$\Delta_{+}\approx 1$ at $z\approx 10$ in order to produce the
observed large scale structure of matter distribution in the
universe. This result only weakly depends on the existence of
dark energy. It is now almost universally accepted that the
primordial density perturbations have been created during the
epoch of very early inflation. Unfortunately so far a
sufficiently detailed model of inflation has not yet been
proposed. Our results show that the constrains placed on the
mechanism of generation of the primordial density perturbations
by the observed large scale structure distribution of baryonic
and dark matter only weakly depend on the existence of dark
energy.

\section*{Acknowledgments}

While working on this paper one of us, MD, has been supported in
part by the Polish State Committee for Scientific Research grant
Nr. 1-P03D-014-26.

\renewcommand{\theequation}{\Alph{section}.\arabic{equation}}
\setcounter{section}{1}
\setcounter{equation}{0}

\section*{Appendix A}

A second order linear differential equation with four regular
singular points $z=\{0,1,d,\infty\}$ in its canonical form is
known as the Heune equation~\cite{Heune,Ronveux,Whittaker&Watson,Snow}, it has the form
    \begin{equation}
f''(\xi)+
    \left(
\frac{\gamma}{\xi}+\frac{\delta}{\xi-1}+\frac{\epsilon}{\xi-d}
    \right)
f'(\xi)+
\frac{\alpha\,\beta \xi-q}{\xi(\xi-1)(\xi-d)}f(\xi)=0,
    \label{eq:A01}
    \end{equation}
where $\alpha$, $\beta$, $\gamma$, $\delta$, $\epsilon$, $d$ and
$q$ are constants. The Heune equation is a natural generalization of the
hypergeometric equation. The Heune equation and its solutions are
determined by the  Riemann $P$-symbol\arraycolsep.25em
    \begin{equation} P
    \left\{
        \begin{array}{ccccc}
    0&1&d&\infty&\\
    0&0&0&\alpha&;\xi\\
    1-\gamma&1-\delta&1-\epsilon&\beta&
        \end{array}
    \right\}.
    \label{eq:A02}
    \end{equation}\arraycolsep.25em
However this symbol does not fully determine the Heune equation
and its solutions because the equation  contains an additional
parameter $q$. The parameters
$\alpha,~\beta,~\gamma,~\delta,~\epsilon$ are not independent and
they are constrained by the Fuchs invariant
    \begin{equation}
1+\alpha+\beta=\gamma+\delta+\epsilon.
    \label{eq:A03}
    \end{equation}

The solution of  the equation (\ref{eq:A01}), which is regular at $\xi=0$
is known as the Heune function and it is defined by
the~series~\cite{Whittaker&Watson}
    \begin{equation}
F(d,q;\alpha,\beta,\gamma,\delta;\xi)=1+\sum\limits_{n=1}^{\infty}
\frac{c_n}{n!(\gamma)_n\mbox{$d^n$}}\xi^n,
    \label{eq:A04}
    \end{equation}
where  $(\gamma)_n\equiv\gamma(\gamma+1)\cdots (\gamma+n-1)$ is
the Pochhammera symbol defined for~$\gamma\neq0,-1,-2\ldots$ The
coefficients $c_n$ of the series (\ref{eq:A04}) are determined by
a recurrence relation of the third order
    \begin{equation}
c_n=
\left\{(n - 1)\left[\alpha + \beta - \delta + n - 1 +
(\gamma + \delta + n - 2)d\right] + q\right\}c_{n - 1} - (\alpha + n - 2)(\beta + n - 2)
(\gamma + n - 2)(n-1)d\, c_{n-2},
    \label{eq:A05}
    \end{equation}
for $n\geq 2$, with the initial condition
    \begin{equation}
c_0=1,~~c_1=q\,.
        \label{eq:A06}
    \end{equation}
The convergence radius of this series (\ref{eq:A04}) is
determined by the distance from the point $\xi=0$ to the nearest
singular point (the radius of convergence $R=1$ for $|d|>1$ or
$R=|d|$ for $|d|<1$). The second
independent solution near the point $\xi=0$ with the exponent
$1-\gamma$ ($\gamma\neq1$) has the form
    \begin{equation}
\xi^{1-\gamma}F(d,q_1;1+\alpha-\gamma,1+\beta-\gamma,2-\gamma,\delta;\xi),
    \label{eq:A07}
    \end{equation}
where
$q_1=q+(1-\gamma)(1+\alpha+\beta-\gamma+(d-1)\delta)$.
When $\gamma=1$ both solutions
(\ref{eq:A04}) and~(\ref{eq:A07}) are identical and then the
second independent solution contains a logarithm (see~\cite{Snow})
and it has the form
    \begin{equation}
G(d,q;\alpha,\beta,1,\delta;\xi)=F(d,q;\alpha,\beta,1,\delta;\xi)\ln \xi
+\sum\limits_{n=0}^\infty b_n \xi^n.
    \label{eq:A08}
    \end{equation}
The coefficients $b_n$ of the series (\ref{eq:A08}) are
determined by the recurrence relation
    \begin{eqnarray}
b_n&=&\frac{1}{d n^2}
        \left\{
    \left[
q+(1+\alpha+\beta - \delta +(\delta+1)d+(d+1)(n-2))(n - 1)
    \right]
b_{n-1}
        \right.
\nonumber\\
&&{}
        -
    \left[
\alpha\beta+(\alpha+\beta)(n - 2)+(n - 2)^2
    \right]
b_{n - 2}
-2\frac{dn}{d^n(n!)^2}c_n\nonumber\\
&&
{}+[1+\alpha+\beta-\delta+(\delta+1)d-(d+1)(1-2(n-1))]
\frac{1}{d^{n-1}[(n-1)!]^2}c_{n-1}\nonumber\\
&&{}
-[\alpha+\beta+2(n-2)]\frac{1}{d^{n-2}[(n-2)!]^2}c_{n-2}
        \left.
        \right\}.
        \label{eq:A09}
    \end{eqnarray}
for $n\geq 2$, with the initial conditions
    \begin{eqnarray}
b_0=0,~~b_1=\frac{1}{d}
    \left[
\alpha+\beta+\delta(d-1)-2c_1
    \right]\,.
        \label{eq:A10}
    \end{eqnarray}

\section*{Appendix B}

\renewcommand{\theequation}{\Alph{section}.\arabic{equation}}
\setcounter{section}{2}
\setcounter{equation}{0}

A second order linear differential equation with six regular
singular points $z=\{0,1,d_1,d_2,d_3,\infty\}$ can be written down
in the following canonical form
    \begin{equation}
f''(\xi)+
    \left(
\frac{\gamma}{\xi}+\frac{\delta}{\xi-1}+\frac{\epsilon_1}{\xi-d_1}
+\frac{\epsilon_2}{\xi-d_2}+\frac{\epsilon_3}{\xi-d_3}
    \right)
f'(\xi)+
\frac{\alpha\,\beta \xi^3+q_2\xi^2+q_1\xi-q}{\xi(\xi-1)(\xi-d_1)(\xi-d_2)(\xi-d_3)}f(\xi)=0,
    \label{eq:B01}
    \end{equation}
and it is a generalization of the Heune equation. With the equation
(\ref{eq:B01}) one can associate the Riemann
 $P$-symbol
    \begin{equation}
P
    \left\{
        \begin{array}{ccccccc}
    0&1&d_1&d_2&d_3&\infty&\\
    0&0&0&0&0&\alpha&;\xi\\
    1-\gamma&1-\delta&1-\epsilon_1&1-\epsilon_2&1-\epsilon_3&\beta&
        \end{array}
    \right\}.
    \label{eq:B02}
    \end{equation}
Because the equation (\ref{eq:B01}) contains three additional
parameters $\{q,q_1,q_2\}$ therefore the Riemann $P$-symbol
(\ref{eq:B02}) does not uniquely determine the form of the
differential equation. The Fuchs invariant puts additional constrains
on the parameters that appear in the equation
$\alpha,\beta,\gamma,\delta,\epsilon_1,\epsilon_2,\epsilon_3$ imposing
the relation
    \begin{equation}
1+\alpha+\beta=\gamma+\delta+\epsilon_1+\epsilon_2+\epsilon_3.
    \label{eq:B03}
    \end{equation}

The solution of the equation (\ref{eq:B01}) which is regular at
$\xi=0$ is a function that is a generalization of the series~(\ref{eq:A04}),
it is given by
    \begin{equation}
F(d_1,d_2,d_3,q,q_1,q_2;\alpha,\beta,\gamma,\delta,\epsilon_1,\epsilon_2;\xi)=1+\sum\limits_{n=1}^{\infty}
\frac{c_n}{n!(\gamma)_n\mbox{$d_1^nd_2^nd_3^n$}}\xi^n.
    \label{eq:B05}
    \end{equation}
The coefficients $c_n$ in the series (\ref{eq:B05}) are determined by
a recurrence relation of the fifth order
    \begin{equation}
c_n=
P_nc_{n - 1} + Q_n c_{n-2}+R_n c_{n-3}+S_n c_{n-4},
    \label{eq:B06}
    \end{equation}
where
\begin{eqnarray}
P_n&=&-q + (n-1 ) (d_2 d_3 (\gamma + \epsilon_1+n-2) +
        d_1 (d_3 (\gamma + \epsilon_2+n-2)\nonumber\\
        &&{}+
              d_2 (\gamma + n-2 +
                    d_3 (\gamma + \delta+n-2) + \epsilon_3))),\nonumber\\
Q_n&=&-d_1 d_2 d_3 (n-1) (\gamma+n-2) (q_1 +
      d_3 (n-2) (\gamma + \epsilon_1 + \epsilon_2+n-3) +
      d_2 (n-2) (\gamma + \epsilon_1+n-3  \nonumber\\
      &&{}+
            d_3 (\gamma + \delta + \epsilon_1) + \epsilon_3+n-3) +
      d_1 (n-2) (\gamma + \epsilon_2+ n-3 +
            d_3 (\gamma + \delta + \epsilon_2+n-3) + \epsilon_3\nonumber\\
      &&{}+
            d_2 (\gamma + \delta + \epsilon_3+n-3))),\nonumber\\
R_n&=&d_1^2 d_2^2 d_3^2 (n-2) (n-1) (\gamma+n-3) (
      \gamma+n-2) (12 d_3 +( 1 + d_3)(n^2+(\gamma+\epsilon_1+\epsilon_2-7)n) - \gamma q_2\nonumber\\
      &&{} - 3 d_3(\gamma+\delta+ \epsilon_1+ \epsilon_2) + n \epsilon_3+ d_3\delta n +
      d_2 (n-3) (\gamma + \delta + \epsilon_1 + \epsilon_3+n-4) \nonumber\\
      &&{} + d_1 (n-3) (\gamma + \delta + \epsilon_2 + \epsilon_3+n-4) -
      3 (\gamma + \epsilon_1 + \epsilon_2 + \epsilon_3-4)),\nonumber\\
S_n&=&-d_1^3 d_2^3 d_3^3 (n-3) (n-2) (n-1) (\gamma+n-4) (\gamma+
      n-3) (\gamma+n-2) (\alpha \beta -
      4 (\gamma + \delta + \epsilon_1 + \epsilon_2 + \epsilon_3-5) \nonumber\\
      &&{}+
       n (\gamma + \delta + \epsilon_1 + \epsilon_2 + \epsilon_3+n-2)),
       \label{eq:B07}
\end{eqnarray}
for $n\geq 4$, with initial conditions
    \begin{eqnarray}
c_0&=&1,\nonumber\\
c_1&=&q,\nonumber\\
c_2&=&q^2+
    \left[\gamma
(d_2d_3+
d_1(d_2+d_3+d_2d_3))+\delta d_1d_2d_3+
d_2d_3\epsilon_1+d_1d_3\epsilon_2+d_1d_2\epsilon_3
    \right]
q
-\gamma\,d_1d_2d_3\,q_1,\nonumber\\
c_3&=&q^3+[(2+3\gamma)(d_1d_2+d_1d_3+d_2d_3+d_1d_2d_3)+3d_1d_2d_3
+3(d_2d_3\epsilon_1+d_1d_3\epsilon_2+d_1d_2\epsilon_3)]q^2\nonumber\\
&&{}+[2d_2^2 d_3^2(\gamma + \epsilon_1)(1 + \gamma + \epsilon_1) +
    2 d_1^2(d_3^2(\gamma+\epsilon_2)(1 + \gamma+\epsilon_2)+
          d_2 d_3((1 + d_3) \gamma^2 +
                2 \epsilon_2\ (d_3\delta+\epsilon_3)\nonumber\\
&&{}                    +\gamma(1 + \epsilon_2 +
                      d_3(1 + \delta+\epsilon_2)+\epsilon_3)) +
          d_2^2((1 + d_3 + d_3^2) \gamma^2 +
                d_3^2\delta(1 + \delta)+\epsilon_3 +
                2d_3\delta \epsilon_3 + \epsilon_3^2\nonumber\\
&&{}                     + \gamma
(1 + 2\epsilon_3 + d_3(1 + d_3 + \delta + 2d_3\delta+\epsilon_3)))) +
    d_1d_2d_3(-(2 + 3\gamma)q_1 +
          2(d_3\gamma(1 + \gamma+\epsilon_1)\nonumber\\
&&{}            + d_3(\gamma + 2 \epsilon_1)\epsilon_2 +
                d_2((1 + d_3)\gamma^2 +
                      2\epsilon_1d_3\delta + \epsilon_3) +\gamma(1 + \epsilon_1 +
                            d_3(1 + \delta+\epsilon_1)+\epsilon_3))))]q\nonumber\\
&&{}-2\gamma[((\delta+1) d_1^2d_2^2d_3^2+d_1 d_2^2d_3^2 + d_1^2d_2 d_3^2
+d_1^2 d_2^2d_3+d_1d_2^2d_3^2\epsilon_1+d_1^2d_2d_3^2\epsilon_2
         +d_1^2d_2^2d_3\epsilon_3)q_1+d_1^2 d_2^2 d_3^2q_2]\nonumber\\
&&{}-2\gamma^2[(d_1^2 d_2^2d_3 + d_1^2d_2 d_3^2 +d_1 d_2^2d_3^2
         + d_1^2d_2^2d_3^2) q_1 +d_1^2 d_2^2 d_3^2q_2
         ].\nonumber\\
        \label{eq:B08}
    \end{eqnarray}
The convergence radius of the series (\ref{eq:B05}) is
determined by the distance between the point $\xi=0$ and the
nearest singular point (the radius of convergence $R=\min\{1,|d_1|,|d_2|,|d_3|\}$).
The second independent solution near
the point $\xi=0$ with the exponent $1-\gamma$ ($\gamma\neq1$)
has the form
    \begin{equation}
\xi^{1-\gamma}F(d_1,d_2,d_3,\hat{q},\hat{q}_1,\hat{q}_2;
1+\alpha-\gamma,1+\beta-\gamma,2-\gamma,\delta,\epsilon_1,\epsilon_2;\xi),
    \label{eq:B09}
    \end{equation}
where $\hat q=q + (1 - \gamma) (d_2 d_3 \epsilon_1 +
        d_1 (d_2 d_3 \delta + d_3 \epsilon_2 + d_2 \epsilon_3))$,
        $\hat{q}_1=q_1 + (1 - \gamma) (d_3 (\epsilon_1 + \epsilon_2) +
        d_2 (\epsilon_1 + d_3 (\delta + \epsilon_1) + \epsilon_3) +
        d_1 (\epsilon_2 + d_3 (\delta + \epsilon_2) + \epsilon_3 +
              d_2 (\delta + \epsilon_3)))$,
        $\hat{q}_2=q_2 - (1 - \gamma) (d_3 \delta + \epsilon_1 +
        d_3 \epsilon_1 + \epsilon_2 + d_3 \epsilon_2 + \epsilon_3 +
        d_2 (\delta + \epsilon_1 + \epsilon_3) +
        d_1 (\delta + \epsilon_2 + \epsilon_3))$.
When
$\gamma=1$ both solutions (\ref{eq:B05}) and (\ref{eq:B08}) are
identical and then the second independent solution contains a
        logarithmic term and is given by
    \begin{equation}
G(d_1,d_2,d_3,q,q_1,q_2;\alpha,\beta,1,\delta,\epsilon_1,\epsilon_2;\xi)
=F(d_1,d_2,d_3,q,q_1,q_2;\alpha,\beta,1,\delta,\epsilon_1,\epsilon_2;\xi)\ln \xi
+\sum\limits_{n=0}^\infty b_n \xi^n.
    \label{eq:B10}
    \end{equation}
The coefficients $b_n$ of the series (\ref{eq:B09}) are determined by
the following recurrence relation
    \begin{equation}
b_n=
P_n^b b_{n - 1}+ Q_n^b b_{n-2}+R_n^b b_{n-3}+S_n^b b_{n-4}+
P_n^c c_{n}+ Q_n^c c_{n-1}+R_n^c c_{n-2}+S_n^c c_{n-3}+T_n^c c_{n-4},
        \label{eq:B11}
    \end{equation}
where
    \begin{eqnarray}
P_n^b&=&\frac{1}{n^2d_1d_2d_3}[q+d_2d_3(n-1)(n-1+\epsilon_1)
+d_1(n-1)(d_3(n-1\epsilon_2)d_2(n-1+\delta)+\epsilon_3))],\nonumber\\
Q_n^b&=&-\frac{1}{n^2d_1d_2d_3}[q_1+d_1(n-2)(n-2+d_2(n-2+\delta+\epsilon_3)
+d_3(n-2+\delta+\epsilon_2)+\epsilon_2+\epsilon_3)\nonumber\\
&&\null+d_2(n-2)(n-2+d_3(n-2+\delta+\epsilon_1)+\epsilon_1+\epsilon_3)
+d_3(n-2)(n-2+\epsilon_1+\epsilon_2)],\nonumber\\
R_n^b&=&-\frac{1}{n^2d_1d_2d_3}[q_2+(n-3)(d_1(1+\delta+\epsilon_2+\epsilon_3)
+d_2(1+\delta+\epsilon_1+\epsilon_3)+d_3(1+\delta+\epsilon_1+\epsilon_2)\nonumber\\
&&\null-(n-4)(n-3)(1+d_1+d_2+d_3)+\epsilon_3)],\nonumber\\
S_n^b&=&-\frac{1}{n^2d_1d_2d_3}[\alpha\beta+(n-4)(n-4+\delta+\epsilon_1+\epsilon_2+\epsilon_3)],\nonumber\\
P_n^c&=&-\frac{2}{n(n!)^2(d_1d_2d_3)^{n}},\nonumber\\
Q_n^c&=&-\frac{1}{[(n-1)!n]^2(d_1d_2d_3)^{n}}[d_1(d_3(2n-2+\epsilon_2)+d_2(2n-2+d_3(2n-2+\delta)+\epsilon_3))+d_2d_3(2n-2+\epsilon_1)],\nonumber\\
R_n^c&=&-\frac{1}{[(n-2)!n]^2(d_1d_2d_3)^{n-1}}[d_1(2n-4+\epsilon_2+\epsilon_3+d_3(2n-4+\delta+\epsilon_2)+d_2(2n-4+\delta+\epsilon_3))\nonumber\\
&&\null+d_2(2n-4+\epsilon_1+\epsilon_3+d_3(2n-4+\delta+\epsilon_1))+d_3(2n-4+\epsilon_1+\epsilon_2)],\nonumber\\
S_n^c&=&-\frac{1}{[(n-3)!n]^2(d_1d_2d_3)^{n-2}}[d_1(2n-6+\delta+\epsilon_2+\epsilon_3)
+d_2(2n-6+\delta+\epsilon_1+\epsilon_3)\nonumber\\
&&\null+d_3(2n-6+\delta+\epsilon_1+\epsilon_2)
+2n-6+\epsilon_1+\epsilon_2+\epsilon_3],\nonumber\\
T_n^c&=&-\frac{1}{[(n-4)!n]^2(d_1d_2d_3)^{n-3}}[2n-8+\delta+\epsilon_1+\epsilon_2+\epsilon_3],
        \label{eq:B12}
    \end{eqnarray}
for $n\geq 4$, with the initial conditions
    \begin{eqnarray}
b_0&=&0,\nonumber\\
b_1&=&\frac{1}{d_1d_2d_3}
    \left[
d_2d_3\epsilon_1+d_1(d_3\epsilon_2+d_2(d_3\delta+\epsilon_3))-2c_1
    \right],\nonumber\\
b_2&=&
\frac{1}{4d_1^2d_2^2d_3^2}
    \left\{
    [d_2d_3\epsilon_1+d_1(d_3\epsilon_2+d_2(d_3\delta+\epsilon_3))]q
    +d_2d_3\epsilon_1(d_2d_3(1+\epsilon_1)
    +2d_1d_3\epsilon_2+2d_1d_2(d_3\delta+\epsilon_3))
    \right.\nonumber\\
&&\left.
    +d_1^2(d_3^2\epsilon_2(1+\epsilon_2)+2d_2d_3\epsilon_2(d_3\delta+\epsilon_3)
    +d_2^2(d_3^2\delta(1+\delta)+(1+2d_3\delta+\epsilon_3)\epsilon_3))
    \right.\nonumber\\
&&\left.
-[2q+d_2d_3\epsilon_1+d_1(d_2d_3\delta+d_3\epsilon_2+d_2\epsilon_3)]c_1-c_2
    \right\},\nonumber\\
b_3&=&
\frac{1}{108d_1^3d_2^3d_3^3}
    \left\{3[d_2d_3\epsilon_1+d_1(d_3\epsilon_2+d_2(d_3\delta+\epsilon_3))]q^2+
    3[-d_2^2d_3^2\epsilon_1(5+3\epsilon_1)-2d_1d_2d_3(d_3(2\epsilon_1+2\epsilon_2
    +3\epsilon_1\epsilon_2)\right.\nonumber\\
    &&\left.\null+d_2(2d_3\delta+2\epsilon_1 +2d_3\epsilon_1+3d_3\delta\epsilon_1
    +2\epsilon_3+2\epsilon_1\epsilon_3))
    -d_1^2(d_3^2\epsilon_2(5+3\epsilon_2)+2d_2d_3(2\epsilon_2
    +d_3(2\delta+2\epsilon_2+3\delta\epsilon_2) \right.\nonumber\\
    &&\left.\null+2\epsilon_3
    +3\epsilon_2\epsilon_3)+d_2^2(d_3^2\delta(5+3\delta)
    +\epsilon_3(5+3\epsilon_3)+d_3(4\delta+4\epsilon_3+6\delta\epsilon_3)))]q
    \right.\nonumber\\
&&\left.\null-12d_1d_2d_3[d_2d_3\epsilon_1+d_1(d_2\epsilon_2+d_2(d_3\delta+\epsilon_3))]q_1
+6(d_2^3d_3^3\epsilon_1(2+3\epsilon_1+\epsilon_1^2)
+3d_1^2d_2d_3^3\epsilon_1\epsilon_2(1+\epsilon_2)\right.\nonumber\\
&&\left.\null+6d_1^2d_2^2d_3^2\epsilon_1\epsilon_2(d_3\delta+\epsilon_3)
+3d_1^2d_2^3d_3\epsilon_1(d_3^3\delta(1+\delta)+2d_3\delta\epsilon_3+\epsilon_3(1+\epsilon_3)) \right.\nonumber\\
&&\left.\null+3d_1d_2^2d_3^2\epsilon_1(1+\epsilon_1)(d_3\epsilon_2+d_2(d_3\delta+\epsilon_3))
+d_1^3(d_3^3\epsilon_2(2+3\epsilon_2+\epsilon_2^2)+3d_2d_3^2\epsilon_2(1+\epsilon_2)(d_3\delta+\epsilon_3)\right.\nonumber\\
&&\left.\null+3d_2^2d_3\epsilon_2(d_3^2\delta(1+\delta)+2d_3\delta\epsilon_3+\epsilon_3(1+\epsilon_3))
+d_2^3(d_3^3\delta(2+3\delta+\delta^2)+3d_3\delta(1+\delta)\epsilon_3+3d_3\delta\epsilon_3(1+\epsilon_3)\right.\nonumber\\
&&\left.\null+\epsilon_3(2+3\epsilon_3+\epsilon_3^2))))
\right.\nonumber\\
&&\left.\null+[-6q^2+3(d_2d_3(8+5\epsilon_1)+d_1d_3(8+5\epsilon_2)+
d_1d_2(8+8d_3+5d_3\delta+5\epsilon_3))q\right.\nonumber\\
&&\left.\null+3(-2d_2^2d_3^2\epsilon_1(2+\epsilon_1)-4d_1d_2d_3(-2q_1+d_3\epsilon_1\epsilon_2)
-4d_1d_2^2d_3\epsilon_1(d_3\delta+\epsilon_3) -2d_1^2(d_3^2\epsilon_2(2+\epsilon_2)
\right.\nonumber\\
&&\left.\null+2d_2d_3\epsilon_2(d_3\delta+\epsilon_3)+d_2^2(d_3^2\delta(2+\delta)+2d_3\delta\epsilon_3+(2+\epsilon_3)\epsilon_3))))]c_1\right.\nonumber\\
&&\left.\null-3[q+d_2d_3\epsilon_1+d_1(d_2d_3\delta+d_3\epsilon_2+d_2\epsilon_3)]c_2-2c_3
    \right\}.\nonumber\\
        \label{eq:B13}
    \end{eqnarray}

\renewcommand{\theequation}{\Alph{section}.\arabic{equation}}
\setcounter{section}{3}
\setcounter{equation}{0}

\section*{Appendix C}

To evaluate the integral
    \begin{eqnarray}
\int\frac{x^{3\slash 2}}{(\Omega_{\rm m}+\Omega_{\rm
k}x+\Omega_{\Lambda}x^3)^{3\slash 2}}dx\,,
    \label{eq:C01}
    \end{eqnarray}
it is necessary to resolve the third order polynomial
    \begin{eqnarray}
\Omega_{\rm m}+\Omega_{\rm k}x+\Omega_{\Lambda}x^3=0,
    \label{eq:C02}
    \end{eqnarray}
into elementary prime factors. The number of real roots of this
polynomial~(\ref{eq:C02}) is determined by the sign of its discriminant
    \begin{eqnarray}
{\cal D}=\frac{4\Omega_{\rm k}^3+27\Omega_{\rm m}^2\Omega_{\Lambda}}{108\Omega_{\Lambda}^3}.
    \label{eq:C03}
    \end{eqnarray}

There are three possibilities:

\litem{({\bf a})} ${\cal D}=0$. This occurs when
$\Omega_{\Lambda}=-\frac{4\Omega_{\rm k}^3}{27\Omega_{\rm m}^2}$
and then the polynomial~(\ref{eq:C02}) can be represented as
    \begin{eqnarray}
\Omega_{\rm m}+\Omega_{\rm k}x+\Omega_{\Lambda}x^3=
\frac{1}{27\Omega_{\rm m}^2}(3\Omega_{\rm m}-\Omega_{\rm k}x)(3\Omega_{\rm m}+2\Omega_{\rm k}x)^2=0.
    \label{eq:C04}
    \end{eqnarray}
In this case the integral~(\ref{eq:C01}) can be expressed by
elementary functions and we get\\
{{\large$\ast$}} (O) models (Fig.~\ref{fig:4}) for~$\Omega_{\rm k}>0$ and
$x\in\left(0,3\Omega_{\rm m}/\Omega_{\rm k}\right)$
    \begin{eqnarray}
\frac{\sqrt{3} \Omega_{\rm m}\sqrt{x}\,(9 \Omega_{\rm m}^2+9 \Omega_{\rm k} \Omega_{\rm m}x+8 \Omega_{\rm k}^2x^2)}
{2 \Omega_{\rm k}^2\sqrt{3 \Omega_{\rm m}- \Omega_{\rm k}x}\,(3 \Omega_{\rm m}+2 \Omega_{\rm k}x)^2}
-\frac{\Omega_{\rm m}}{2\Omega_{\rm k}^{5\slash{2}}}
\arctan\sqrt{\frac{3 \Omega_{\rm k}x}{3 \Omega_{\rm m}- \Omega_{\rm k}x}}\,,
    \label{eq:C05a}
    \end{eqnarray}
and\\
{{\large$\ast$}} (O) models (Fig.~\ref{fig:2}) for $\Omega_{\rm k}<0$ and
$x\in \left(0,-\frac{3}{2}\Omega_{\rm m}/\Omega_{\rm k}\right)$
    \begin{eqnarray}
\frac{\sqrt{3} \Omega_{\rm m}\sqrt{x}\,(9 \Omega_{\rm m}^2+9 \Omega_{\rm k} \Omega_{\rm m}x+8 \Omega_{\rm k}^2x^2)}
{2 \Omega_{\rm k}^2\sqrt{3 \Omega_{\rm m}- \Omega_{\rm k}x}\,(3 \Omega_{\rm m}+2 \Omega_{\rm k}x)^2}
-\frac{\Omega_{\rm m}}{2\Omega_{\rm k}^2\sqrt{-\Omega_{\rm k}}}
\arctanh\sqrt{-\frac{3 \Omega_{\rm k}x}{3 \Omega_{\rm m}- \Omega_{\rm k}x}}\,.
    \label{eq:C05b}
    \end{eqnarray}

\litem{({\bf b})} ${\cal D}>0$. This happens when
$\Omega_{\Lambda}>0$ and $4\Omega_{\rm k}^3+27\Omega_{\rm
m}^2\Omega_{\Lambda}>0$ or $\Omega_{\Lambda}<0$ and
$4\Omega_{\rm k}^3+27\Omega_{\rm m}^2\Omega_{\Lambda}<0$. In
this case the polynomial (C.2) has one real root and two
complex ones. In this case the polynomial (\ref{eq:C02}) can
be written as
    \begin{eqnarray}
\Omega_{\rm m}+\Omega_{\rm k}x+\Omega_{\Lambda}x^3=
\Omega_{\Lambda}(x -\alpha)(q + p x + x^2)=0,
    \label{eq:C06}
    \end{eqnarray}
where $\alpha=\left(-\frac{\Omega_{\rm m}}{2\Omega_\Lambda}+\sqrt{{\cal D}}\right)^{1\slash3}
+\left(-\frac{\Omega_{\rm m}}{2\Omega_\Lambda}-\sqrt{{\cal D}}\right)^{1\slash3}$,
$q=\frac{\Omega_{\rm k}}{\Omega_\Lambda}+\alpha^2$ and $p=\alpha$.
In this case the integral (\ref{eq:C01}) is given by\\
{{\large$\ast$}} (H) models (Fig.~\ref{fig:2}) for $\Omega_\Lambda>0$ and $4\Omega_{\rm k}^3
+27\Omega_\Lambda\Omega_{\rm m}^2>0$, and $x\in(0,\infty)$
    \begin{eqnarray}
&&{}-\frac{2(-9q^2\alpha^2+3q\alpha(q-\alpha^2)x+2(q-\alpha^2)^2x^2)}{\Omega_\Lambda^{3/2}
(-4q+\alpha^2 )( q+2\alpha^2)^2\sqrt{x(x-\alpha)(q+\alpha x+x^2)}}\nonumber\\
&&{}+\frac{18q^{3/2}\alpha^2}{\Omega_\Lambda^{3/2}(4q-\alpha^4)(q+2\alpha^2)^2}
\frac{\sqrt{(x-\alpha)(q+\alpha x+x^2)}}{\sqrt{x}(\sqrt{q}(x-\alpha^2)\sqrt{q+2\alpha^2}x)}\nonumber\\
&&{}-\frac{18q^{5/4}\alpha}{\Omega_\Lambda^{3/2}(4q-\alpha^4)(q+2\alpha^2)^{7/4}}
    E\left(\arccos(u),k\right)
-\frac{3q^{3/4}\alpha(-3\sqrt{q}+\sqrt{q+2\alpha^2})}{\Omega_\Lambda^{3/2}(4q-\alpha^4)(q+2
\alpha^2)^{7/4}}
    F\left(\arccos(u),k\right).\hspace*{10mm}
    \label{eq:C07b}
    \end{eqnarray}
where $u=\frac{\sqrt{q}(x-\alpha)-x\sqrt{q+2\alpha^2}}{\sqrt{q}(x-\alpha)+x\sqrt{q+2\alpha^2}}$
and $k=\left[\frac12+\frac{2q+\alpha^2}{4\sqrt{q(q+2\alpha^2)}}\right]^{1/2}$,\\
and by\\
{{\large$\ast$}} (O) models (Fig.~\ref{fig:4}) for $\Omega_\Lambda<0$ and $4\Omega_{\rm k}^3
+27\Omega_\Lambda\Omega_{\rm m}^2<0$, and $x\in(0,\alpha)$
    \begin{eqnarray}
&&-\frac{2((2q+\alpha^2)x+3q\alpha)\sqrt{x}}{(-\Omega_\Lambda)^{3/2}
(4q-\alpha^2)(q+2\alpha^2)\sqrt{(\alpha-x)(q+\alpha x+x^2)}}\nonumber\\
&&+\frac{18q\alpha^2}{(-\Omega_\Lambda^{3/2})(4q-\alpha^4)(q+2\alpha^2)^{3/2}}
\frac{\sqrt{q+\alpha x+x^2}}{\sqrt{\alpha-x}(\sqrt{q}(\alpha-x)+\sqrt{q+2\alpha^2}x)}\nonumber\\
&&-\frac{18q^{5/4}\alpha}{(-\Omega_\Lambda)^{3/2}(4q-\alpha^2)(q+2\alpha^2)^{7/4}}
    E\left(\arccos(u),k\right)
+\frac{3q^{3/4}\alpha(3\sqrt{q}+\sqrt{q+2\alpha^2})}{(-\Omega_\Lambda)^{3/2}(4q-\alpha^2)(q+2\alpha^2)^{7/4}}
    F\left(\arccos(u),k\right),\nonumber\\
    \label{eq:C07c}
    \end{eqnarray}
where $u=\displaystyle{\frac{\sqrt{q}(\alpha-x)-x\sqrt{q+2\alpha^2}}{\sqrt{q}(\alpha-x)+x\sqrt{q+2\alpha^2}}}$
and $k=\displaystyle{\left[\frac12-\frac{2q+\alpha^2}{4\sqrt{q(q+2\alpha^2)}}\right]^{1/2}}$.

\litem{({\bf c})} ${\cal D}<0$. This case occurs when  $\Omega_{\Lambda}>0$ and
$4\Omega_{\rm k}^3+27\Omega_{\rm m}^2\Omega_{\Lambda}<0$ or $\Omega_{\Lambda}<0$ and
$4\Omega_{\rm k}^3+27\Omega_{\rm m}^2\Omega_{\Lambda}>0$.
Then the polynomial (\ref{eq:C02}) can be written as
    \begin{eqnarray}
\Omega_{\rm m}+\Omega_{\rm k}x+\Omega_{\Lambda}x^3=\Omega_{\Lambda}(x - \alpha)(x-\beta)(x-\gamma)=0.
    \label{eq:C10}
    \end{eqnarray}
When $\Omega_{\Lambda}<0$ and $4\Omega_{\rm k}^3+27\Omega_{\rm m}^2\Omega_{\Lambda}>0$ we have
\begin{eqnarray*}
\a=2\sqrt{\frac{\Omega_{\rm k}}{-3\Omega_{\Lambda}}}\cos\left[\frac{\phi}{3}\right],\quad
\b=-2\sqrt{\frac{\Omega_{\rm k}}{-3\Omega_{\Lambda}}}\cos\left[\frac{\phi}{3}+\frac{\pi}{3}\right],\quad
\g=-2\sqrt{\frac{\Omega_{\rm k}}{-3\Omega_{\Lambda}}}\cos\left[\frac{\phi}{3}-\frac{\pi}{3}\right],
    \label{eq:C10a}
    \end{eqnarray*}
where $\phi=\arccos\left[\frac{\Omega_{\rm m}}{-2\Omega_{\Lambda}}
\left(\frac{\Omega_{\rm k}}{-3\Omega_{\Lambda}}\right)^{-3/2}\right]$.\\
When $\Omega_{\Lambda}>0$ and $4\Omega_{\rm k}^3+27\Omega_{\rm m}^2\Omega_{\Lambda}<0$ we have
    \begin{eqnarray*}
\a=2\sqrt{\frac{-\Omega_{\rm k}}{3\Omega_{\Lambda}}}\cos\left[\frac{\phi}{3}-\frac{\pi}{3}\right],\quad
\b=2\sqrt{\frac{-\Omega_{\rm k}}{3\Omega_{\Lambda}}}\cos\left[\frac{\phi}{3}+\frac{\pi}{3}\right],\quad
\g=-2\sqrt{\frac{-\Omega_{\rm k}}{3\Omega_{\Lambda}}}\cos\left[\frac{\phi}{3}\right],
    \label{eq:C10b}
    \end{eqnarray*}
where $\phi=\arccos\left[\frac{\Omega_{\rm m}}{2\Omega_{\Lambda}}
\left(\frac{-\Omega_{\rm k}}{3\Omega_{\Lambda}}\right)^{-3/2}\right]$.\\
{{\large$\ast$}} This case corresponds to (O) models (Fig.~\ref{fig:4}) for
$\Omega_{\Lambda}<0$ and $4\Omega_{\rm k}^3+27\Omega_{\rm m}^2\Omega_{\Lambda}>0$.
The roots of the polynomial (\ref{eq:C10}) have been chosen so that
$\a>0>\b>\g$,  and $x\in(0,\a)$. In this case the integral (\ref{eq:C01}) can be
expressed by the elliptic integrals
     \begin{eqnarray}
&&\frac{2(\a\g(\a-2\b+\g)+(\a(\b-2\g)+\b\g)x)\sqrt{x}}
{(\a-\b)(\b-\g)(\a-\g)^2\sqrt{(\a-x)(x-\b)(x-\g)}(-\Omega_\Lambda)^{3/2}}\nonumber\\
&&+\frac{2\g[(\b+\g)(\a^2+\b\g)+\a(\a^2-6\b\g+\g^2)]}
{(\a-\b)(\a-\g)^2(\b-\g)^2\sqrt{-\g(\a-\b)}(-\Omega_\Lambda)^{3/2}} E\left(\arcsin\sqrt{u},k\right)\nonumber\\
&&+\frac{2\b\g(-2\a+\b+\g)}{(\a-\b)(\a-\g)(\b-\g)^2\sqrt{-\g(\a-\b)}(-\Omega_\Lambda)^{3/2}}
        F\left(\arcsin\sqrt{u},k\right),
    \label{eq:C13a}
    \end{eqnarray}
where $u=\displaystyle{\frac{(\a-\b)x}{\a(x-\b)}}$ and $k=\displaystyle{\left[\frac{\a(\g-\b)}{(\a-\b)\g}\right]^{1/2}}$.\\
{{\large$\ast$}} (B) models (Fig.~\ref{fig:2}) for $\Omega_{\Lambda}>0$ and
$4\Omega_{\rm k}^3+27\Omega_{\rm m}^2\Omega_{\Lambda}<0$.
The roots of the polynomial (\ref{eq:C10}) have been ordered so that $\a>\b>0>\g$,
and $x\in(\a,\infty)$.
The integral (\ref{eq:C01}) is given in terms of the elliptic integrals
     \begin{eqnarray}
&&\frac{2\b[(\b+\g)(\a^2+\b\g)+\a(\b^2-6\b\g+\g^2)]}
{[(\a-\b)(\a-\g)(\b-\g)]^2\Omega_\Lambda^{3/2}}\frac{\sqrt{(x-\a)(x-\g)}}{\sqrt{x(x-\b)}}\nonumber\\
&&-\frac{2\a\b\g((\b+\g)(\a^2+\b\g)+\a(\b^2-6\b\g+\g^2))}
{[(\a-\b)(\a-\g)(\b-\g)]^2\Omega_\Lambda^{3/2}\sqrt{x(x-\a)(x-\b)(x-\g)}}\nonumber\\
&&+\frac{2[\b^2\g^2(-2\a+\b+\g)+\a^3(\b^2+\g^2)+\a^2(\b+\g)(\b^2-3\b\g+\g^2)]x}{[(\a-\b)(\a-\g)(\b-\g)]^2\Omega_\Lambda^{3/2}\sqrt{x(x-\a)(x-\b)(x-\g)}}\nonumber\\
&&-\frac{4[\a^2(\b^2-\b\g+\g^2)+\b\g(\b\g-\a(\b+\g))]x^2}
{[(\a-\b)(\a-\g)(\b-\g)]^2\Omega_\Lambda^{3/2}\sqrt{x(x-\a)(x-\b)(x-\g)}}\nonumber\\
&&-\frac{2\sqrt{\a}[(\b+\g)(\a^2+\b\g)+\a(\b^2-6\b\g+\g^2)]}
{(\a-\b)^2(\a-\g)^2(\b-\g)^{3/2}\Omega_\Lambda^{3/2}} E\left(\arcsin\sqrt{u},k\right)\nonumber\\
&&+\frac{2\sqrt{\a}\g(\a-2\b+\g)}{(\a-\b)(\a-\g)^2(\b-\g)^{3/2}\Omega_\Lambda^{3/2}}
        F\left(\arcsin\sqrt{u},k\right),
    \label{eq:C13b}
    \end{eqnarray}
where $u=\displaystyle{\frac{(\b-\g)(x-\a)}{(\a-\g)(x-\b)}}$ and $k=\displaystyle{\left[\frac{\b(\a-\g)}{\a(\b-\g)}\right]^{1/2}}$.\\
{{\large$\ast$}} (O) models (Fig.~\ref{fig:2}) for $\Omega_{\Lambda}>0$ and
$4\Omega_{\rm k}^3+27\Omega_{\rm m}^2\Omega_{\Lambda}<0$.  The roots of the polynomial (\ref{eq:C10})
have been ordered so that $\a>\b>0>\g$ and~$x\in(0,\b)$.  In this case the integral~(\ref{eq:C01})
can be also expressed by the elliptic integrals
     \begin{eqnarray}
&&-\frac{2\a\b\g[(\b+\g)(\a^2+\b\g)+\a(\b^2-6\b\g+\g^2)]}
{[(\a-\b)(\a-\g)(\b-\g)]^2\Omega_\Lambda^{3/2}\sqrt{x(\a-x)(\b-x)(x-\g)}}\nonumber\\
&&+\frac{2[\b^2\g^2(-2\a+\b+\g)+\a^3(\b^2+\g^2)+\a^2(\b+\g)(\b^2-3\b\g+\g^2)]x}{[(\a-\b)(\a-\g)(\b-\g)]^2\Omega_\Lambda^{3/2}\sqrt{x(\a-x)(\b-x)(x-\g)}}\nonumber\\
&&-\frac{4[\a^2(\b^2-\b\g+\g^2)+\b\g(\b\g-\a(\b+\g))]x^2}
{[(\a-\b)(\a-\g)(\b-\g)]^2\Omega_\Lambda^{3/2}\sqrt{x(\a-x)(\b-x)(x-\g)}}\nonumber\\
&&-\frac{2\a[\a^2(\b+\g)(\a^2+\b\g)+\a(\b^2-6\b\g+\g^2)]}{[(\a-\b)(\a-\g)(\b-\g)]^{2}\Omega_\Lambda^{3/2}}
        \frac{\sqrt{(\b-x)(x-\g)}}{\sqrt{x(\a-x)}}\nonumber\\
&&+\frac{2\sqrt{\a}[(\b+\g)(\a^2+\b\g)+\a(\b^2-6\b\g+\g^2)]}{(\a-\b)^2(\a-\g)^2(\b-\g)^{3/2}\Omega_\Lambda^{3/2}}
        E\left(\arcsin\sqrt{u_1},k\right)\nonumber\\
&&+\frac{2\g[(\b+\g)(\a^2+\b\g)+\a(\b^2-6\b\g+\g^2)]}{\sqrt{\a}(\a-\b)(\a-\g)^2(\b-\g)^{5/2}\Omega_\Lambda^{3/2}}
        F\left(\arcsin\sqrt{u_1},k\right)\nonumber\\
&&-\frac{2\b\g(-2\a+\b+\g)}{\sqrt{\a}(\a-\b)(\a-\g)^2(\b-\g)^{5/2}\Omega_\Lambda^{3/2}}
        F\left(\arcsin\sqrt{u_2},k\right),
    \label{eq:C13c}
    \end{eqnarray}
where $u_1=\displaystyle{\frac{\a(\b-x)}{\b(\a-x)}}$, $u_2=\displaystyle{\frac{(\b-\g)x}{\b(x-\g)}}$
and $k=\left[\displaystyle{\frac{\b(\a-\g)}{\a(\b-\g)}}\right]^{1/2}$.

\renewcommand{\theequation}{\Alph{section}.\arabic{equation}}
\setcounter{section}{4}
\setcounter{equation}{0}

\section*{Appendix D}

The standard classification of the Friedman cosmological models
describing evolution of the universe that is filled in with
dust, radiation and non zero cosmological constant is based
on a technique used already by Friedman who studied the
behavior of the curve ${\dot x}(t)=0$ (the so called zero
velocity curve). The zero velocity curve corresponds to the
isoclines in higher dimensional dynamical systems. (However in
our case the equation ${\dot x}(t)=0$ does not determine
positions of the fixed points. In the Friedman equation ${\dot
x}(t)$ appears squared and therefore in the equation ${\dot
x}(t)=F(x)$ the right hand side does not define
diffeomorphic mapping at $x=0$.

\begin{figure}[th]
\centerline{\psfig{file=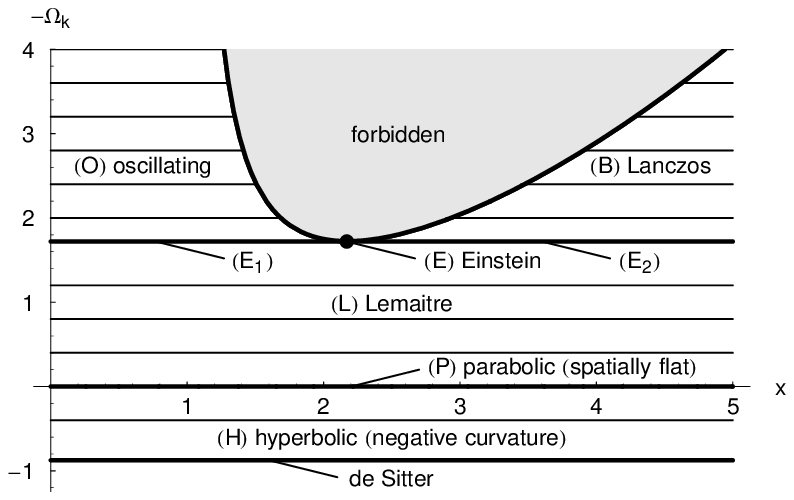,width=10cm,angle=0}}
\caption{The $\{x,\Omega_{\rm k}\}$ plane for
$\Omega_\Lambda>0$.}
    \label{fig:2}
\centerline{\psfig{file=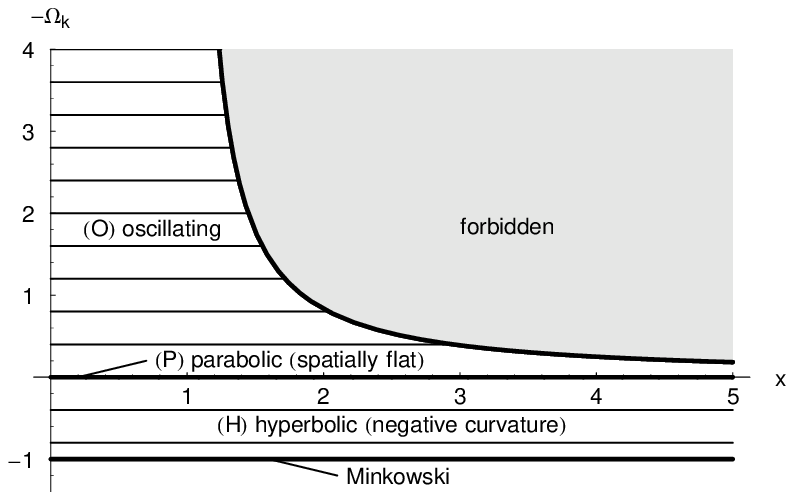,width=10cm,angle=0}}
\caption{The $\{x,\Omega_{\rm k}\}$ plane for
$\Omega_\Lambda=0$.}
    \label{fig:3}
    \end{figure}

\begin{figure}[th]
\centerline{\psfig{file=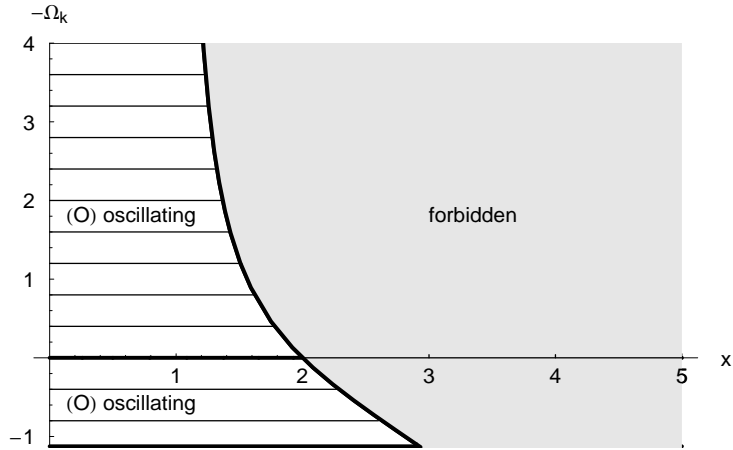,width=10cm,angle=0}}
\caption{The $\{x,\Omega_{\rm k}\}$ plane for
$\Omega_\Lambda<0$.}
    \label{fig:4}
    \end{figure}

In the classical papers (\cite{Friedman,Robertson}, see
also~\cite{Stabell&Refsdal,Harrison,Ehlers&Rindler}) the zero
velocity curve is determined on the surface
$\{x,\Omega_\Lambda\}$. We use a more transparent classification
on the surface  $\{x,\Omega_{\rm k}\}$ on which it is easy to
identify solutions that correspond to phase trajectories in the
classification of Belinsky and Khalatnikov~\cite{Bel&Khal}. From
the Friedman equation~(\ref{eq:3}) and~(\ref{eq:4}),
(\ref{eq:6}), (\ref{eq:7}) and the definitions of the Omega
parameters~(\ref{eq:00}) we get
     \begin{equation}
\frac{1}{H_0^2}\frac{{\dot x}^2(t)}{x^2(t)}
=\frac{\Omega_{\rm r}}{x^4(t)}+\frac{\Omega_{\rm m}}{x^3(t)}
+\frac{\Omega_{\rm k}}{x^2(t)}+\Omega_\Lambda,
    \label{eq:D00}
    \end{equation}
where $x(t)=\frac{a(t)}{a_0}$. In this case the equation that
determines the isocline
has the form
     \begin{equation}
-\Omega_{\rm k}=\frac{1}{x-1}+(1+x+x^2)\Omega_\Lambda-\frac{\Omega_{\rm r}}{x}\,.
     \label{eq:D01}
    \end{equation}
The isocline has a vertical asymptote at $x=1$. From the point
of view of cosmological applications  the most interesting is
the region where $\Omega_{\rm m}\geq 0$ and $\Omega_{\rm r}\geq
0$. In Figs.~\ref{fig:2},~\ref{fig:3} and~\ref{fig:4} the non physical region where ${\dot x}$
is complex is shaded and denoted as ``forbidden''. In Figs.~\ref{fig:2},~\ref{fig:3}
and~\ref{fig:4} we represent respectively the allowed parameter space for
$\Lambda>0$, $\Lambda=0$ and $\Lambda<0$ restricted to the half
plane representing universes with positive mass-energy density.

In each of this figures the allowed cosmological models are
represented by horizontal segments or half lines in the non
shaded region that is bounded by the zero velocity curve. The
models (O) are oscillating. For such models the universe starts
its evolution form singularity and the universe expands until
${\dot x}=0$ and then contracts to the final singularity. The
models (B) are possible only for $\Lambda > 0$, they represent
the non singular Lanczos models. Such models describe the
universe that starts expanding from a finite volume and expands
to infinity. There is a possibility of bouncing when the
universe initially contracts having a finite volume larger than
the minimal possible for the given set of parameters. The half
line that is tangent to the zero velocity curve represents two
models ${\mbox{E}_1}$ and ${\mbox{E}_2}$ both approaching the static Einstein
model (E) (the point of intersection) correspondingly in the
forward and backward direction in time. The horizontal half line
below the critical one represents the Lema\^{\i}tre models.
The $x$ axis represents the flat models (parabolic (P) models in
the Ehlers-Rindler classification~\cite{Ehlers&Rindler}) and the
half lines below represent universes with negative spacial
curvature (hyperbolic (H) models in the Ehlers-Rindler
classification~\cite{Ehlers&Rindler}). When $\Lambda=0$ the
space of solutions contains only models (O), (P), (H) and the
Minkowski space. When $\Lambda<0$ only oscillating models (O)
are possible.

Let us note that the zero velocity curve ${\dot x}=0$ does not
cross the line $x=0$. The crossing condition
    \begin{equation}
\Omega_\Lambda x^4+\Omega_{\rm m}x +\Omega_{\rm r}=0
    \label{eq:D02}
    \end{equation}
does not have real solutions when all
$\{\Omega_\Lambda,\Omega_{\rm m},\Omega_{\rm r}\}$ are positive.
It means that the cosmological models with negative spacial
curvature and positive cosmological constant independently on
the values of the other cosmological parameters always start
their evolution from the initial singularity and expand to
infinity. We consider those models in our analysis of evolution
of density perturbations.

\def \Journal#1#2#3#4{(#1)~{#2}~{#3},~#4}
\def \AA{{\em Astronomy Astroph.}}
\def \ApJ{{\em Astroph.~J.}}
\def \ASS{{\em Astroph. Space Sci.}}
\def \CQG{{\em Class. Quantum Grav.}}
\def \GRG{{\em Gen. Rel. Grav.}}
\def \MA{{\em Math. Ann.}}
\def \MNRAS{{\em Mon. Not. R. Astron. Soc.}}
\def \PAZh{{\em Pis'ma Astron. Zh.}}
\def \RMP{{\em Rev. of Mod. Phys.}}
\def \ZhETPh{{\em ZhETPh}}
\def \Science{{\em Science}}
\def \ZfP{{\em Zeits. f\"ur Phys.}}
\def \PRD{{\em Phys.~Rev.}~D}
\def \PTP{{\em Prog.~Theor.~Phys.}}

%\newpage

\edoc